# Multi-k magnetic structure and large anomalous Hall effect in candidate magnetic Weyl semimetal NdAlGe

C. Dhital[1*,†], R. L. Dally[2,†], R. Ruvalcaba[3], R. Gonzalez-Hernandez[4], J. Guerrero-Sanchez[3], H. B. Cao[5], Q. Zhang[5], W. Tian[5], Y. Wu[5], M. D. Frontzek[5], S. K. Karna[6,7], A. Meads[1], B. Wilson[1], R. Chapai[7], D. Graf[8], J. Bacsa[9], R. Jin[7,10], and J.F. DiTusa[7,11]

[1] Department of Physics, Kennesaw State University, Marietta, GA, 30060, USA

[2] NIST Center for Neutron Research, National Institute of Standards and Technology, Gaithersburg, MD 20899-6102, USA

[3] Centro de Nanociencias y Nanotecnología, Universidad Nacional Autónoma de México, Ensenada, BC 22860, México

[4] Departamento de Física y Geociencias, Universidad del Norte, Barranquilla, Colombia.

[5] Neutron Scattering Division, Oak Ridge National Laboratory, Oak Ridge, Tennessee 37831, USA

[6] Department of Physics, Prairie View A&M University, Prairie View, TX, 77446, USA

[7] Department of Physics, Louisiana State University, Baton Rouge, LA, 70803, USA

[8] National High Magnetic Field Laboratory, Tallahassee, FL, 32310, USA

[9] X-ray Crystallography Center, Department of Chemistry, Emory University, Atlanta, GA, 30322, USA

[10] Center for Experimental Nanoscale Physics, Department of Physics and Astronomy, University of South Carolina, Columbia, SC, 290208, USA

[11] Department of Physics, The Purdue School of Science, IUPUI, Indianapolis, Indiana 46202, USA

## Abstract

The magnetic structure, magnetoresistance, and Hall effect of the non-centrosymmetric magnetic semimetal NdAlGe are investigated revealing an unusual magnetic state and anomalous transport properties that are associated with the electronic structure of this compound. The magnetization and magnetoresistance measurements are both highly anisotropic and indicate an Ising-like magnetic system. The magnetic structure is complex in that it involves two magnetic ordering vectors including an incommensurate spin density wave and commensurate ferrimagnetic state in zero field. We have discovered a large anomalous Hall conductivity that reaches ≈ 430 $\Omega^{-1}$cm$^{-1}$ implying that it originates from an intrinsic Berry curvature effect stemming from Weyl nodes found in the electronic structure. These electronic structure calculations indicate the presence of nested Fermi surface pockets with nesting wavevectors similar to the measured magnetic ordering wavevector and the presence of Weyl nodes in proximity to the Fermi surface. We associate the incommensurate magnetic structure with the large anomalous Hall response to be the result of the combination of Fermi surface nesting and the Berry curvature associated with Weyl nodes.

*Corresponding author: cdhital@kennesaw.edu
† These authors contributed equally to this work.

## 1. Introduction

Weyl semimetals are characterized by linearly dispersing electronic bands near the Fermi surface and are classified into different types based on the degeneracies and distribution of nodes in momentum space [1,2]. The emergence of Weyl semimetal phases requires either broken space inversion symmetry or broken time-reversal symmetry [1–4]. When both symmetries are broken as in non-centrosymmetric magnetic compounds, then there is a possibility of tuning Weyl nodes using the coupling between topology and magnetism [5,6]. Materials having heavy rare earth elements are suitable for the study of intertwined electronic band topology and magnetism due to their penchant for long-range magnetic ordering, strong spin-orbit coupling, and low carrier density semi-metallic behavior. Recent studies on compounds belonging to RAlX (R=Rare earth, X=Si, Ge) have shown promising results suitable for the investigation of coupling between electronic topology and magnetism [7–10].

The RAlX family of materials has shown interesting magnetic orders and exciting electrical transport properties. The list of these properties includes Lorentz violating type II Weyl Fermions in LaAlGe [11], Fermi arcs and large anomalous Hall effect in PrAlGe [9,12,13], Kramers nodal lines in SmAlSi [14], singular magnetoresistance in CeAlGe [10,15], an intrinsic to extrinsic crossover in the anomalous Hall conductivity PrAlGe$_{1-x}$Si$_x$ [16], and anisotropic anomalous Hall effect, magnetostriction effects, and Fermi arcs in CeAlSi [17,18]. The magnetic behaviors and magnetic structures also exhibit interesting variations including topologically non-trivial magnetic orders [8,9,12,17]. All these observations indicate that this family of materials can be a productive place to search for exciting electronic and magnetic properties arising from the interplay of electronic topology and magnetism. With this motivation, we have investigated NdAlGe using magnetometry, electrical transport, neutron diffraction, and electronic structure calculations.

There are some recent magnetization and magnetotransport studies [19–22] performed on NdAlGe that indicate some of its magnetic properties are similar to sister compound NdAlSi [8], which hosts incommensurate magnetism mediated by Weyl fermions. The similarities between NdAlGe and NdAlSi include the non-centrosymmetric crystal structure, metamagnetic transition in-field, small value of saturation moment ($\approx$ 2.8 $\mu_B$) compared to expected free ion ($Nd^{3+}$) moment value ($\approx$3.6 $\mu_B$), and highly anisotropic magnetism. Despite many similarities, some differences were also reported in the same previous works. The main difference is the absence of an anomalous Hall effect in NdAlSi compared to the large anomalous Hall response in NdAlGe [19]. The other difference was the observation of two successive magnetic transitions in NdAlSi [8] but a single magnetic transition in NdAlGe [19–21]. Therefore, it is natural to ask why there are some key differences in the magnetic and anomalous Hall behavior between these two isostructural compounds with the same magnetic ion. The first step towards answering such questions is to understand the magnetic structures of these compounds. Although the magnetic structure of NdAlSi has been solved in a previous study [8], the magnetic structure of NdAlGe is still missing. In this work, we have solved the ground state magnetic structure of NdAlGe, and performed a systematic investigation correlating its magneto transport, and anomalous Hall response with its magnetic structure and electronic topology by employing neutron diffraction, magnetization, magnetotransport measurements, and electronic structure calculations. Our results indicate that NdAlGe crystallizes in a non-centrosymmetric tetragonal structure with a space group ($I4_1md$) identical to other members of the RAlX family [21,23–25]. We observed two successive magnetic transitions in *dc* magnetic susceptibility, heat capacity, resistivity, and the neutron order parameter at $T_{IC}$ = 6.3 K and $T_C$ = 4.9 K. These observations are similar to two transitions observed in NdAlSi [8] but differ from the recent observations [19,20] of single magnetic transition in NdAlGe. Our orientation-dependent magnetization and magnetoresistance measurements indicate a highly anisotropic Ising-like magnetic system in agreement with previous studies [19,20]. There is a non-monotonic variation of magnetoresistance which reaches up to 15% ($T$ = 0.4 K, 18 T) for $H // c$ and 5% ($T$ = 0.4 K, 18T) for $H \perp c$. It has an unusually high anomalous Hall response most likely originating from the intrinsic contribution from Berry curvature related to Weyl nodes near the Fermi surface. Our neutron diffraction measurements indicate an incommensurate ($\delta \approx 0.006$ r.l.u) ($T_C < T < T_{IC}$) to commensurate ($\delta = 0$) ($T < T_C$) magnetic order transition leading to a ground state defined by two magnetic wavevectors: $\mathbf{k_0} = 0$ and $\mathbf{k_1} = (\frac{1}{3}, \frac{1}{3}, 0)$. The ordered moment at 1.5 K is 3.03(9) $\mu_B$/Nd with a net ferromagnetic component of 1.01(3) $\mu_B$/Nd, consistent with the bulk magnetization measurements. The Fermi surface shows nested hole pockets located at the magnetic ordering wavevector $\pm (\frac{1}{3}, \frac{1}{3}, 0)$ and Weyl nodes near the Fermi level. Thus, the magnetic behavior and magnetic structure of NdAlGe are similar to NdAlSi except for the fact that the quantum oscillations have not been observed in NdAlGe. The absence of quantum oscillations is most probably caused by the higher disorder as indicated by the residual resistivity ratio, RRR = 2.5, and the low temperature resistivity, $\rho_{xx}$ = 23 $\mu\Omega$-cm at $T$ = 2 K for crystals grown thus far. Our data, which demonstrate both the formation of incommensurate spin density wave and large anomalous Hall response, indicate that Weyl fermions are involved in forming the magnetic order through the cooperative interplay between nested itinerant fermions and the RKKY interaction between local moments.

## 2. Experimental Details

Single crystals of NdAlGe were obtained using the flux method with excess Al serving as the flux. The elements were loaded in an alumina crucible in the ratio Nd: Ge: Al (1:2:20) and sealed inside a quartz tube under partial Ar pressure. The mixture was heated to 1100 $^0$C in 4 hrs, homogenized at 1100 $^0$C for 2 hrs then cooled to 700 $^0$C at the rate of 8 $^0$C/hr. The solution was subsequently centrifuged to obtain single crystals. The excess Al flux was removed using a NaOH: H$_2$O solution. The phase purity and crystallinity of the resulting crystals were probed using both powder and single crystal x-ray diffraction. Multiple single-crystal magnetic neutron diffraction experiments were performed at High Flux Isotope Reactor (HFIR) facility in Oak Ridge National Laboratory (ORNL) using the single crystal diffractometer HB-3A in four circle and two-axis mode, the Wide-Angle Neutron Diffractometer (WAND$^2$, HB-2C), the HB-1A triple-axis spectrometer, and the HB-2A powder diffractometer. The experiment performed at HB-3A employed the four-circle mode with a closed-cycle refrigerator having a minimum temperature of 4.8 K. Subsequently, we employed this instrument in two-axis mode with a liquid helium cryostat having a minimum temperature of 1.5 K. Both experiments performed at HB-3A used a wavelength of 1.54 Å. Neutron experiments performed on the HB-2C WAND$^2$ diffractometer at Oak Ridge National Laboratory (ORNL) used a liquid helium cryostat with a minimum temperature of 1.5 K and a Ge (113) monochromator with a wavelength of 1.48 Å. Neutron experiments performed on the HB-1A triple-axis spectrometer at ORNL utilized a liquid helium cryostat with a minimum temperature of 1.5 K and a wavelength of 2.37 Å. Both a PG(002) monochromator and analyzer were used and Söller collimators before the monochromator, before the sample, after the sample, and before the detector was 40′ − 40′ − 40′ − 80′ , respectively.

The sample was oriented in the (H, K, 0) scattering plane. The neutron powder diffraction at HB-2A employed a neutron beam with a wavelength of 2.41 Å, defined by a Ge(113) crystal monochromator. Neutron diffraction data were fit using Rietveld refinement with the program, FullProf [26].

The magnetization measurements were carried out using Quantum Design Magnetic Property Measurement System (MPMS) at Louisiana State University (LSU) and National High Magnetic Field Laboratory (NHMFL) (SCM-5). The high-field magnetization measurements were carried out using cell 8 (35 T) at NHMFL. The magnetoresistance measurements were performed using four probe methods on a Quantum Design Physical Property Measurement System (PPMS) and an 18 T superconducting magnet (SCM-2) at NHMFL. Four electrodes were mounted on a single crystal sample using silver epoxies. The current was applied along the $a$-axis and the sample rotation angle was measured from the $c$-axis such that $\theta = 0°$ corresponds to $H // c$ and $\theta = 90°$ corresponds to $H // a \perp c$. Heat capacity measurements were performed at LSU using a PPMS. The magnetic field values are expressed in kOe (10 kOe = 1 T) units throughout this manuscript.

For the electronic structure calculations, we employed the plane wave pseudopotential implementation of DFT contained in the Vienna ab initio Simulation Package (VASP) [27–29]. The PBE-GGA [30] description of the exchange-correlation energy was employed. Relaxation of the atomic positions was carried out using the lattice parameters values: $a = b = 4.2270$ Å and $c = 14.6051$ Å . The convergence criterion for the total energy was set to $10^{-4}$ eV, paying special attention to keeping the I4$_1$md spatial symmetry in the system. A cutoff energy of 400 eV was set for the plane wave basis. A $k$-points grid of 5 × 5 × 5 was used to discretize the first Brillouin zone, tripling it for the band structure calculations. This grid was generated using the Monkhorst-Pack method [31] and a Methfessel-Paxton [32] smearing of the second order with a width of 0.05 eV. We took into consideration the spin-orbit coupling (SOC) and Hubbard U corrections to account for the relativistic effects and the highly correlated *f*-orbitals of the Nd atoms, respectively. The Hubbard corrective functional was set with an effective on-site Coulomb interaction of *U* = 6 eV on the *f*-orbital of the Nd atoms as used in previous studies [8,17] to improve the description of electronic correlation and localization. Calculations were performed using different permutations of these corrections (DFT, DFT+SOC, DFT+U, DFT+SOC+U), but only the results of the latter two are reported. The crystal structures, data processing, and graphs in this paper were generated using VESTA [33], VASPKIT [34], and Matplotlib [35] software, respectively. The Weyl nodes were determined using the Wannier Tools package [36] after an accurate interpolation of the FM band structure at the Fermi level with the Wannier 90 code [37].

Throughout the manuscript, error bars and uncertainties represent plus and minus one standard deviation.

## 3. Results

### 3. a Crystal Structure

Due to possible variations in the crystal structure of chemistry such as $NdGe_{2-x}Al_x$ and previous reports [38–40] which suggest a stoichiometry-dependent crystal structure, we performed crystal structure analysis employing different probes and different batches of samples. Powder x-ray diffraction was used to check for phase purity and single crystal x-ray diffraction, powder neutron diffraction, and single crystal neutron diffraction were employed to analyze the crystal structure. The results from single crystal x-ray diffraction, powder neutron diffraction, and single-crystal neutron diffraction are presented in Table I and appendix A. Both powder and single crystal samples are better refined with a non-centrosymmetric space group $I4_1md$ (#109) compared to centrosymmetric structure $I4_1/amd$ (#141). The non-centrosymmetric space group is further confirmed by the value of Flack parameter being 0.02±0.01. The Flack parameter gives the absolute orientation of non-centrosymmetric crystal. The non-centrosymmetric structure ($I4_1md$) is an ordered structure of LaPtSi type where Ge and Al are ordered whereas the centrosymmetric structure is the disordered structure of type $\alpha$-$ThSi_2$ where Al and Ge occupy the same site randomly with a 50:50 ratio [38]. Instead of a 1:1:1 ratio of Nd: Ge: Al, we also noticed some variations in stoichiometry for samples from different synthesis sets.

**Table I:** Results of crystal structure analysis using single crystal x-ray, neutron powder diffraction, and single crystal neutron diffraction. The difference in the goodness of refinement is presented.

| T = 173 K | T = 20 K (Powder) | T = 20 K (Powder) | T = 4.8 K |
|---|---|---|---|
| Single crystal x-ray | Powder neutron | Powder neutron | Single crystal neutron |
| Non-centrosymmetric | Non-centrosymmetric | Centrosymmetric | Non-centrosymmetric |
| Space group: $I4_1md$ | Space group: $I4_1md$ | Space group: $I4_1/amd$ | Space group: $I4_1md$ |
| a = 4.23076 (15) Å | a = 4.2255(2) Å | a = 4.2274 (1) | a = 4.224 (2) Å |
| c = 14.6364(4) Å | c = 14.6042(6) Å | c = 14.6107 (4) | c = 14.624 (3) Å |
| Z = 4 | Z = 4 | Z = 4 | Reflections: 127 |
| Reflections = 629 | R-factor = 2.06 | R-factor = 6.5 | R-factor = 2.55 |
| R ($F^2$) = 1.22 | RF-factor = 1.62 | RF-factor = 8.3 | RF-factor = 3.77 |
| $R_1$ = 0.027, $wR_2$ = 0.032 | | | |
| Flack parameter = 0.02 ± 0.01 | | | |

The Rietveld fittings are presented in Appendix A. Column II and III represent refinement of same data using two different space groups. The R-factor and RF factor are better for non-centrosymmetric group.

## 3. b Magnetic Properties

The low field magnetic properties of NdAlGe are presented in Fig. 1. The magnetic properties are highly anisotropic with $(\chi_c/\chi_{ab})_{dc} = 83.6$ at $T= 2$ K as evident from Fig. 1a. The variation of *ac* susceptibility (Fig. 1b) shows a broad transition centered around 4.5 K, whereas the variation of *dc* susceptibilities for $H \;//\; c$ (Fig. 1c) and $H \;//\; a$ (Fig. 1d) with temperature indicate two possible magnetic transitions between $T_{IC} = 6.3$ K and $T_C = 4.9$ K. These transitions are more evident in heat capacity measurement presented in Fig 3.

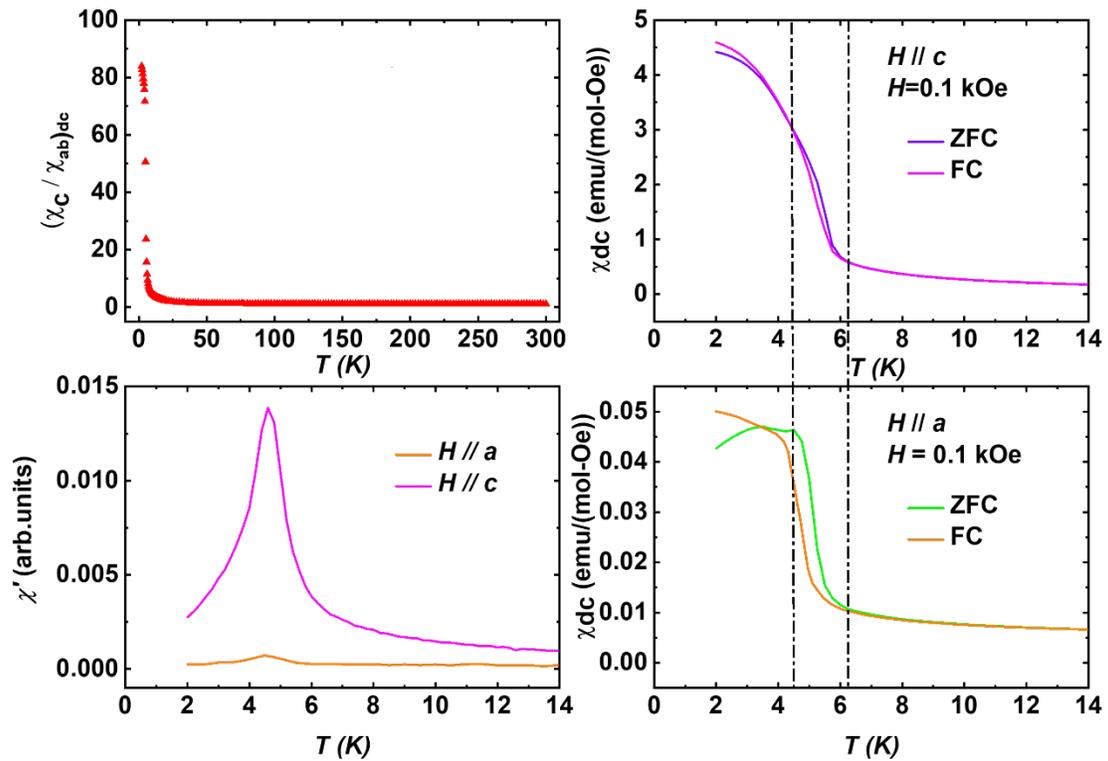

**Fig. 1** ac and dc magnetic susceptibility measurements for NdAlGe. (a) The ratio of *dc* susceptibilities for $H \parallel c$, $\chi_c$, and $H \parallel a$, $\chi_a$, measured at $H = 0.1$ kOe (b) Variation of the real, $\chi'$, and imaginary, $\chi''$, parts of the *ac* magnetic susceptibility with temperature, *T*, for $H_{ac} = 1$ Oe and $f = 20$ Hz. (c) Variation of zero field cooled (ZFC) and field cooled (FC) susceptibilities with *T* for $H \parallel c$ (d) Variation of (ZFC) and (FC) *dc* susceptibilities with temperature for $H \parallel a$ (1 emu/(mol Oe)) = $4\pi \ 10^{-6}$ m³/mol).

For both field orientations ($H \parallel c$ and $H \parallel ab$), the susceptibility follows Curie-Weiss behavior $\chi(T) = \frac{C}{T-\theta}$, resulting in $\mu_{eff} = 3.42(3)$ $\mu_B$/Nd, $\theta = +4.2$ (2) K for $H \parallel c$ and $\mu_{eff} = 3.59(5)$ $\mu_B$/Nd, $\theta = -2.2$ K for $H \parallel ab$. The effective moment $\mu_{eff}$ is slightly less than 3.66 $\mu_B$ as expected for $Nd^{3+}$ ion.

We also investigated the variation of the magnetization, *M*, as a function of the magnetic field, *H*. The results are presented in Fig. 2. We observe two distinct steps in the magnetization as a function of *H* for $T < T_C$. For $H \parallel c$, the magnetization increases linearly and sharply until it reaches a magnetization plateau (shaded region I in Fig. 2a and b) at $M_1 = 0.95(5)$ $\mu_B$ for $H_1 = 1$ kOe. Here, the values of magnetization and field are defined for the center of the plateau and center of transition, respectively. With further increase in field, *M* remains relatively constant (5% change) until the field reaches 28 kOe (at $T = 2$ K). With further increase in the field, there is a second stepwise increase in magnetization. The magnetization increases rapidly until the field reaches 44 kOe, and after that remains relatively constant (2% variation, shaded region II in Fig. 2a and b), reaching a value $M_2 = 2.8$ $\mu_B$ at 70 kOe at $T = 2$ K. Here, the stepwise increase starts at 28 kOe and ends at 44 kOe so we define the midfield $H_2 = 36$ kOe as the second critical field. The value of $H_1$ remains constant between 2 K and 4 K, whereas the value of $H_2$ decreases with increasing temperature. Furthermore, the sharp increase in magnetization below $H_1$ is non-hysteretic (Fig. 2b) indicating the sharp increase is likely not due to domain alignment. The increase in magnetization near $H_2$ displays a small hysteresis (Fig. 2b) at $T = 2$ K $< T_C$ but is non-hysteretic at $T = 5$ K between $T_C$ and $T_{IC}$. Since the value of $M$ is less than that expected for localized $Nd^{3+}$ ion for fields up to 70 kOe, we performed magnetization measurements up to 350 kOe for fields $H // c$ and $H // a$ as shown in Fig. 2c. The value of $H_2$ (115 kOe) for $H // a$ is much larger than the value of $H_2$ (36 kOe) for $H // c$ reflecting a strong Ising-like magnetic anisotropy of the system. We also observed that the magnetization at $T = 1.8$ K and $H = 350$ kOe only reaches up to 2.81 $\mu_B$ for $H // c$ and 2.2 $\mu_B$ for $H // a$. These values are still significantly smaller than the expected free ion ($Nd^{3+}$) moment of 3.66 $\mu_B$. This indicates the importance of itinerant magnetic moments to the magnetic state of this compound. In an itinerant ferromagnet, the small variation of $dc$ magnetization in the field polarized phase generally originates from Pauli susceptibility of itinerant carriers.

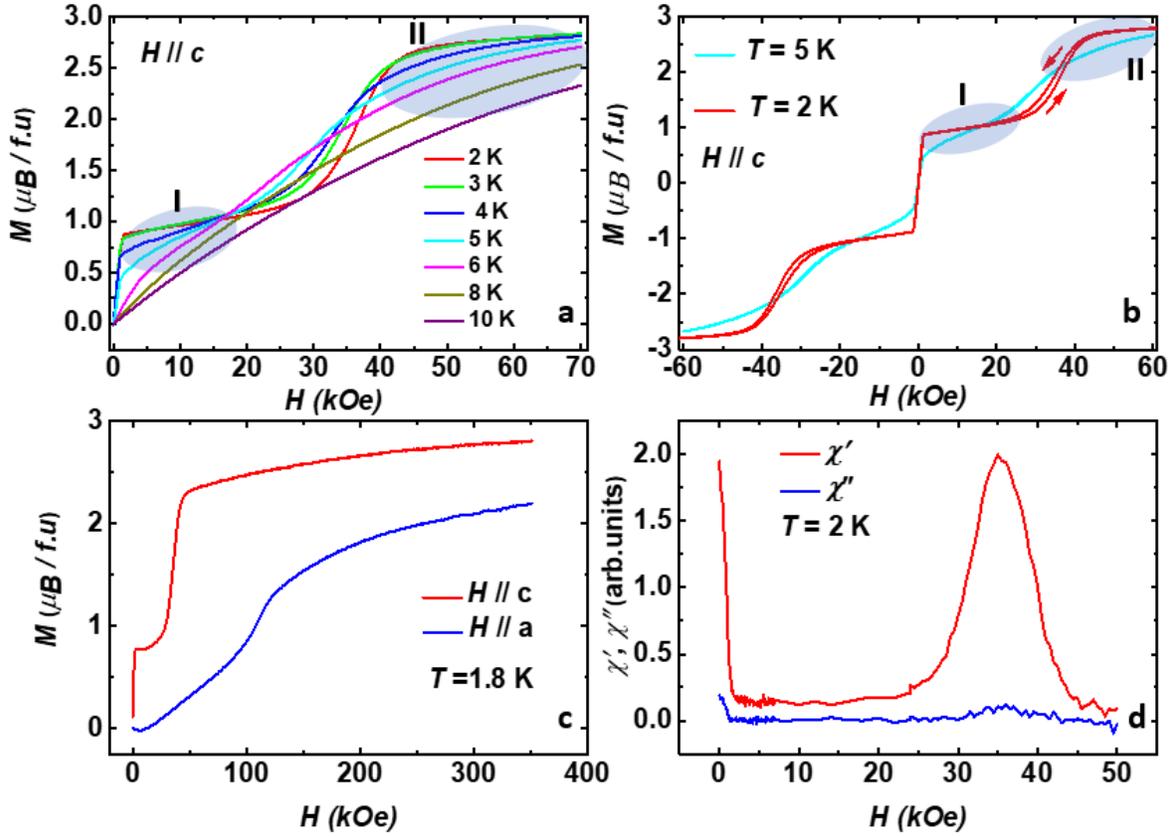

**Fig.2** Magnetization of NdAlGe. (a) Magnetization, *M*, vs magnetic field, *H*, at select temperatures for *H // c* (b) Hysteresis loop at 5 K and 2 K . The red arrows indicate increasing and decreasing fields at 2 K. The two plateau regions in magnetization are shaded and labeled as regions I and II (c) *M* vs *H* for *H // c* and *H* ⊥ c (d) Real part ($\chi'$) and imaginary part ($\chi''$) vs *H* at *T* = 2 K with $H_{ac}$ = 1 Oe and *f* = 20 Hz.

We have also investigated the field dependent dynamic behavior of *M* via ac-susceptibility measurements. Fig. 2d summarizes the results for *H // c*, $H_{ac}$ = 2 Oe and *f* =100 Hz. For $H < H_1$, there is a sharp decrease in $\chi'$ and $\chi''$ followed by a peak near the metamagnetic transition $H_2$ (=36 kOe). These two features correspond to $H_1$ and $H_2$ as defined in *dc* magnetization measurements. The sharp decrease below $H_1$ is an indication of either domain alignment or the presence of non-collinear magnetic structure. The broad peak around $H_2$ is an indication of magnetic structure change due to spin reorientation. Both $\chi'$ and $\chi''$ remain almost flat above 44 kOe when the system fully enters into field polarized phase.

## 3. c Specific Heat Capacity

We also measured the specific heat capacity of NdAlGe as a function of field and temperature. These are summarized in Fig. 3. Fig. 3a presents heat capacity as a function of temperature at select magnetic fields. There is a λ-anomaly near the magnetic transition temperature for fields $H \leq 10$ kOe as well as a Schottky-like anomaly centered around 17 K. A closer look at the region about the critical temperature reveals two distinct peaks (for $H = 0$) at $T_C = 4.9$ K and $T_{IC} = 6.3$ K (Fig 3b) indicating a sequence of two magnetic transitions. These two peaks merge for fields above $H_1$, become broader above $H_2$, and move to a higher temperature at higher magnetic fields indicating the possible mixing of crystal field levels. The zero-field heat capacity is fit with the Debye model [41]($C_L$) with the best fit found for $\theta_D = 240$ K (blue curve in Fig. 3b). The pink curve in Fig.3b represents the magnetic contribution to the specific heat $C_m = C_p - C_L$. The inset in Fig. 3b shows the variation of magnetic entropy: $Sm = \int_0^T \frac{Cm}{T} dT$ as a function of temperature. A total magnetic entropy of Rln2 just above $T_{IC}$ and 0.93R ln (10) is recovered between 2 K and 100 K indicating an $L = 6$, $S = 3/2$, $J=9/2$ ground state of $Nd^{3+}$ similar to NdAlSi [8].

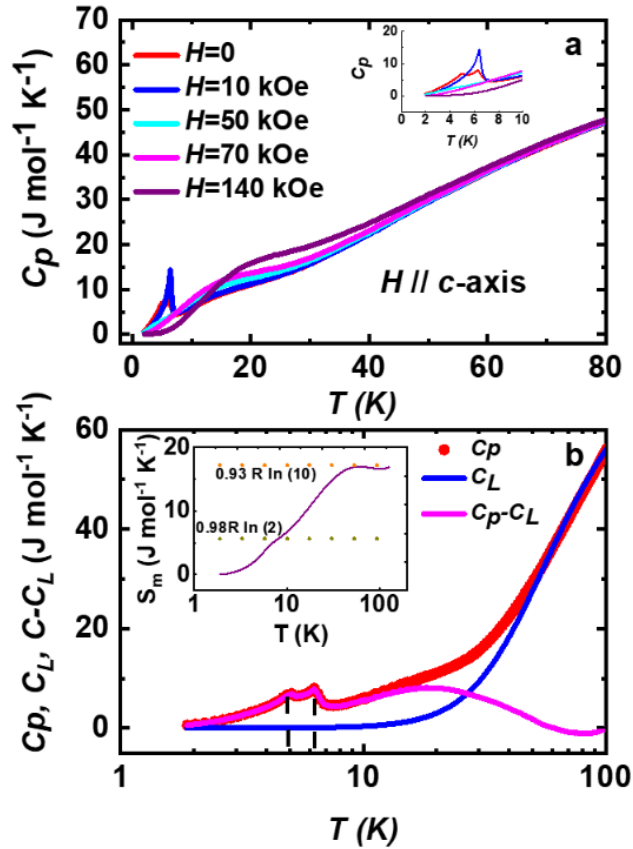

**Fig.3** Specific Heat capacity of NdAlGe (a) Variation of specific heat capacity, $C_P$, with temperature, $T$, at select magnetic fields for $H // c$. Inset shows $C_P$ below 10 K (b) Zero field heat capacity (red), fitted by the Debye model, $C_L$ (blue), magnetic component to specific heat, $C_m = C_p - C_L$, and magnetic entropy, $S_m$ (inset) as a function of temperature, $T$.

### 3.d Magnetoresistance and Hall Effect

We also performed resistivity measurements in magnetic fields up to 180 kOe (18 T) for different orientations of the magnetic field to explore the charge degrees of freedom and the influence of the magnetic ordering. The variation of resistivity with temperature is shown in Fig 4a and in Appendix B. The resistivity, $\rho_{xx}$, decreases monotonically (See Appendix B) from $T = 300$ K to $T_{IC} = 6.3$ K. The resistivity displays anomalies at both $T_{IC} = 6.3$ K and $T_C = 4.9$ K. Such non-monotonic variation can arise either due to a reduction in carrier density due to the formation of incommensurate spin density wave (SDW) state [42,43] or due to enhanced scattering from domain walls or from the critical magnetic fluctuations. Below $T_C$, the resistance drops sharply due to a reduction in spin disorder scattering. Fig. 4b presents the magnetoresistance of NdAlGe at temperatures between 2 K and 10 K for fields up to $H = 90$ kOe for $\theta = 0°$ ($H // c$-axis). For $T < T_C$, the magnetoresistance is positive up to the metamagnetic transition at $H_2$. At $H_2$, the magnetoresistance drops significantly and then increases at higher field. The decrease in resistance at $H_2$ may be ascribed to the formation of a polarized magnetic state and reduction in spin disorder scattering. For $T_C < T < T_{IC}$ (i.e. at 5 K and 6 K), the magnetoresistance decreases sharply around $H = 0$ (shaded area in Fig. 4b) and then increases again for fields above $H_1$. This indicates that the zero-field magnetic structure between $T_C$ and $T_{IC}$ is different than that below $T_C$. To investigate the possibility of observing quantum oscillations and the variation of magnetoresistance above $H_2$, we also performed orientation and field-dependent magnetoresistance measurements up to 180 kOe using the superconducting magnet SCM-2 at NHMFL. The results are presented in Fig. 4c. The magnetoresistance peaks near $H_2$ and increases monotonically for $H > H_2$ reaching up to 15% at $T = 0.4$ K and $H = 180$ kOe for $H // c$ ($\theta = 0°$). A large positive magnetoresistance in the field-polarized magnetic state ($H > H_2$) is rather counterintuitive and cannot be simply described by the Lorentz contribution. This implies the possibility of Fermi surface reconstruction as suggested for NdAlSi [44] and other $f$-electron compounds [45]. Unfortunately, we did not observe quantum oscillations in measurements of the resistivity. The inherent disorder present in our sample (RRR = 2.5) is likely the cause and does not allow a validation of a scenario of changing the Fermi surface topology. We also observed that the value of $H_2$ increases monotonically with $\theta$ at the same time the magnetoresistance decreases with $\theta$. This is expected in a magnetically anisotropic system where the spins are primarily oriented along the $c$-axis.

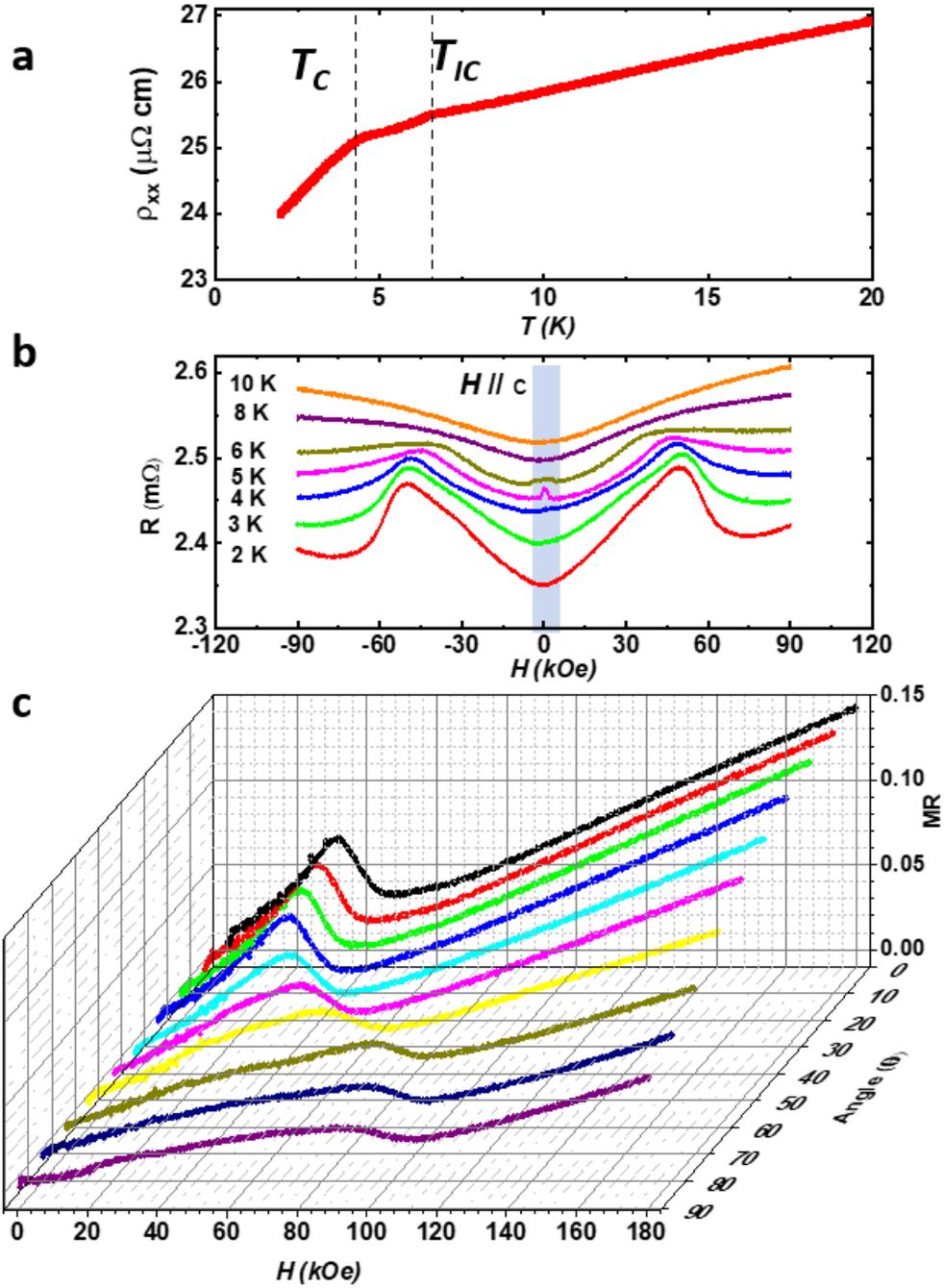

**Fig. 4** Magnetotransport behavior of NdAlGe. (a) Variation of zero field resistivity, $\rho_{xx}$, with temperature, $T$. (b) Variation of resistance, $R$, with the magnetic field, $H$, at different temperatures for $H \parallel c$ (c) Variation of magnetoresistance, $MR$ ($MR = \frac{R(H)-R(0)}{R(0)}$), with the magnetic field, $H$ for different orientations of the magnetic field at $T = 0.4$ K.

We have also measured the Hall effect for NdAlGe with the results are presented in Fig. 5. The symmetrized Hall resistivity, $\rho_{xy}$ plotted as a function of $H$ in Fig. 5a, interestingly exhibits two regions with different slopes for $T \leq T_{IC}$. These two regions are more clear in the inset of Fig.5a and in Appendix C. These two regions correspond to two different plateaus (I and II) observed in magnetization measurements in Fig. 2a and 2b. The data in Fig. 5a can be fit (for $T \leq T_{IC}$) with two straight lines using the relation: $\rho_{xy} = R_0 H + \rho_{xy}^A$ [46,47], where the slope $R_0$ is the ordinary Hall coefficient and the intercept $\rho_{xy}^A$ gives the anomalous Hall resistivity. The anomalous conductivity, $\sigma_{xy}^A$, can then be calculated using the relation: $\sigma_{xy}^A = \frac{\rho_{xy}^A}{(\rho_{xy}^A)^2 + (\rho_{xx})^2}$. From the two different intercepts in $T \leq T_{IC}$, we estimated large values of $\sigma_{xy}^A$ to be $|\sigma_{xy}^{A,\,I}| \approx 430$ $\Omega^{-1}$ cm$^{-1}$ and $|\sigma_{xy}^{A,\,II}| \approx 1030$ $\Omega^{-1}$ cm$^{-1}$, respectively, at $T = 2$ K. These values decrease to $|\sigma_{xy}^{A,\,I}| \approx 170$ $\Omega^{-1}$ cm$^{-1}$ and $|\sigma_{xy}^{A,\,II}| \approx 530$ $\Omega^{-1}$ cm$^{-1}$, respectively, at $T = 6$ K. To take into account the variation of magnetization and resistivity with temperature, we also analyzed the variation of $\frac{\rho_{xy}}{H}$ with $\frac{M}{H}$ at different temperatures. Fig. 5b shows the variation of $\frac{\rho_{xy}}{H}$ with $\frac{M}{H}$ at different temperatures. Each temperature dataset for $T \leq T_{IC}$ in Fig. 5b can be divided into two linear regions (Inset of Fig. 5b) corresponding to two plateaus (I and II) in magnetization data. For $T > T_{IC}$, the straight region corresponding to the lower plateau (region I) disappears and can be fit with a single straight line in the high field region (plateau II). The linear region in $\frac{\rho_{xy}}{H}$ vs $\frac{M}{H}$ can be modelled using the equation for Hall resistivity: $\frac{\rho_{xy}}{H} = R_0 + 4\pi R_S \frac{M}{H}$. In this equation, the intercept, $R_0$, represents the ordinary Hall coefficient which in the simplest model is $1/ne$ where $n$ is the carrier density and $e$ is the electronic charge. The slope, $4\pi R_S$, provides the anomalous Hall coefficient, $R_S$. For $T \leq T_{IC}$, fitting of straight lines in regions I (4 kOe-20 kOe) and II (46 kOe-70 kOe) gives two values of $R_S$, i.e. $R_S^I$ and $R_S^{II}$. The intercept $R_0$ is the same (within error) for the two regions. For $T \geq T_{IC}$, only the region with a slope $R_S^{II}$ survives.

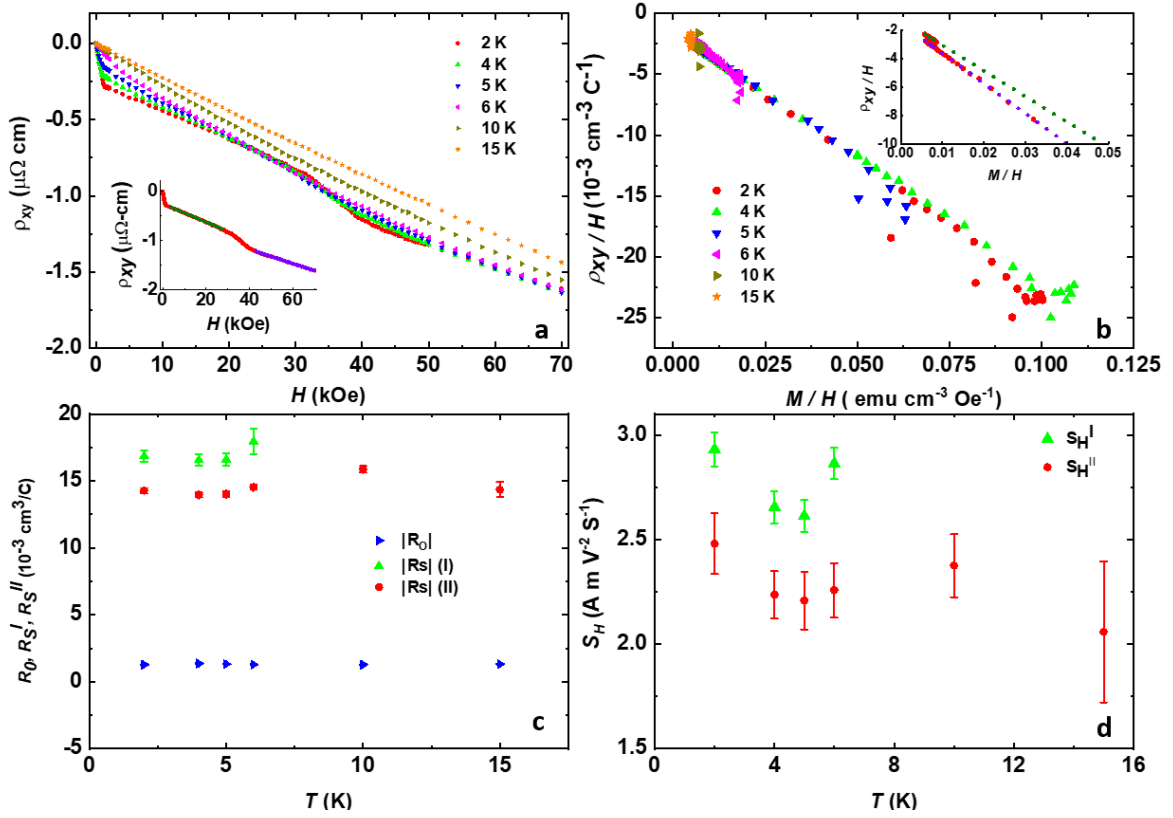

**Fig. 5** Hall effect in NdAlGe. (a) Variation of Hall resistivity, $\rho_{xy}$, with the magnetic field, $H$ at selected temperatures. The Hall resistivity was symmetrized to remove the effect of longitudinal resistivity. The inset shows variation of $\rho_{xy}$ at $T = 2$ K. The $T = 2$ K data is fit with two linear regions (olive and violet) with different intercepts. (b) Variation of $\frac{\rho_{xy}}{H}$ with $\frac{M}{H}$ at select temperatures. Inset of (b) shows data at $T = 2$ K. The data at $T = 2$ K is fit with two linear regions with different slopes (1 emu/(Oe cm$^3$) = 4π). (c) Variation of the magnitude of the ordinary ($R_0$) and anomalous ($R_S^I$ and $R_S^{II}$) Hall coefficients with temperature. (d) Variation of the magnitude of the Hall conductivity parameter, $S_H$, with temperature.

The variation of of $|R_0|$, $|R_S^I|$, and $|R_S^{II}|$ with temperature is plotted in Fig. 5c. It should be noted that the anomalous component, $R_S$, is almost 10 times larger than the ordinary component, $R_0$, comparable to other compounds of this family [9,10,13,16]. The intrinsic or extrinsic contributions to the anomalous Hall effect can be deduced from the temperature dependence of the parameter $S_H$ where $S_H = \frac{R_S}{\rho_{xx}^2}$. Fig. 5d shows the variation of $S_H$ with $T$ for the two different anomalous contributions. In the small range of temperature studied in this work, both $S_H$ values seem relatively independent of the temperature. Given the large values of Hall conductivities and the relative temperature-independent $S_H$, the anomalous Hall effect is likely to originate from the intrinsic Berry curvature-related phenomena [48,49]. We also analyzed the possibility of a topological Hall effect (in region I) and anomalous Hall effect in (in region II). These less likely scenarios are discussed in the discussion and conclusion sections.

## 3. e Neutron diffraction

The measurements of magnetization, heat capacity, and resistivity indicate two successive magnetic transitions at $T_{IC}$ = 6.3 K and $T_C$ = 4.9 K. To solve the magnetic structure of NdAlGe we have performed multiple neutron diffraction experiments. Fig. 6 summarizes the main results and shows the magnetic order parameters (Figs. 6b and c from the HB-1A experiment), the ground state magnetic structure (Fig. 6a), and the goodness-of-fit for the $T$ = 4.8 K refinement taken in four-circle mode on the HB-3A diffractometer (Fig. 6d).

The ground state magnetic structure can be described by the multi-**k** structure with propagation vectors $\mathbf{k}_0 = 0$ and $\mathbf{k}_1 = \left(\frac{1}{3}, \frac{1}{3}, 0\right)$. The order parameters for the two components can be seen in Figs. 6b and c, respectively, where intensity at **Q** = (0,2,0), $\left(\frac{1}{3}, \frac{1}{3}, 0\right)$, and $\left(\frac{2}{3}, \frac{2}{3}, 0\right)$ increases beginning below $T_{IC}$. The program k-SUBGROUPSMAG on the Bilbao Crystallographic Server [50] was used to calculate all the possible magnetic symmetries for a magnetic ordering with propagation vector $\mathbf{k}_1$ and crystallographic space group, $I4_1md$. Here, a ($3a$, $3b$, $c$) supercell is defined and leads to three possible magnetic space groups which allow for a $c$-axis ferromagnetic component alongside the $\left(\frac{1}{3}, \frac{1}{3}, 0\right)$ modulation. The magnetic space group $Fd`d`2$ (#43.227) was found to best fit the data at 4.8 K, just below the commensurate magnetic transition. It should be noted that the other two possible magnetic space groups, $Cc`$ (9.39) and $C2$ (5.13), are subgroups of $Fd`d`2$, which is a k-maximal subgroup and has a higher symmetry.

There are two Nd sites in the magnetic unit cell, $\mathbf{r}_1 = (0,0,0)$ and $\mathbf{r}_2 = \left(\frac{1}{6},\frac{1}{6},\frac{1}{2}\right)$ (these are the positions $(0,0,0)$ and $\left(\frac{1}{2},\frac{1}{2},\frac{1}{2}\right)$ in the nuclear unit cell, respectively); they are antiparallel to one another and oriented along the c-axis, creating an up-up-down and down-down-up motif, respectively, along the **k**-vector direction. The moments were constrained to have equal magnitude, yielding 1.34(2) $\mu_B$/Nd and a net ferromagnetic moment of 0.45(2) $\mu_B$/Nd. Fig. 6d shows the refinement results for 100 $\left(\frac{1}{3},\frac{1}{3},0\right)$-type reflections, with a goodness-of-fit $R_F$-factor = 6.60. The ab-canting (discussed shortly) was zero within error if left to refine. At 1.5 K, using the data taken from HB-3A in two-axis mode, the refinement yields 3.03(9) $\mu_B$/Nd with a net ferromagnetic moment of 1.01(3) $\mu_B$/Nd. This experiment yielded significantly less data within the detector coverage and a goodness-of-fit $R_F$-factor = 9.08.

This purely c-axis magnetic structure does not give rise to magnetic Bragg peak intensity at all positions where we observed weak intensity, e.g. at $\left(\frac{1}{3},\frac{1}{3},0\right)$ shown in Fig. 6c. Intensity at these positions can occur with ab-plane canting, like that observed in NdAlSi [8]. Magnetic space group $Fd`d`2$ allows for canting at the Nd $\mathbf{r}_1$ position, forming double-FM stripes perpendicular to the **k**-vector which alternate direction travelling along the **k**-vector (see Fig. 17 in Appendix D). The symmetry allows for canting along the a-axis, b-axis, or mixed. If we assume the moments are along the [-1,1,0] direction like in NdAlSi, then we can estimate the 1.5 K value using data collected during our HB-1A experiment. There are very few reflections due to the canting that do not also overlap with the strong Bragg peaks stemming from the c-axis up-up-down structure. Data at only six such positions was collected during our experiments, which prevents a full Rietveld refinement; however, we can estimate the canting to be 0.22(1) $\mu_B$ using the available data.

The magnetic structure described above assumes that only one arm-of-the-star for $\mathbf{k} = \left(\frac{1}{3}, \frac{1}{3}, 0\right)$ participates in forming the magnetic structure. It is possible that the whole star forms a multi-k structure, although the neutron diffraction experiments presented here can not distinguish between these two possibilties. For example, a multi-k structure was proposed for the related Weyl semimental compound, CeAlGe, where the multi-k structure retains a higher symmetry than the single-k solution and is topologically non-trivial in a finite field [10]. A more detailed discussion is included in Appendix D.

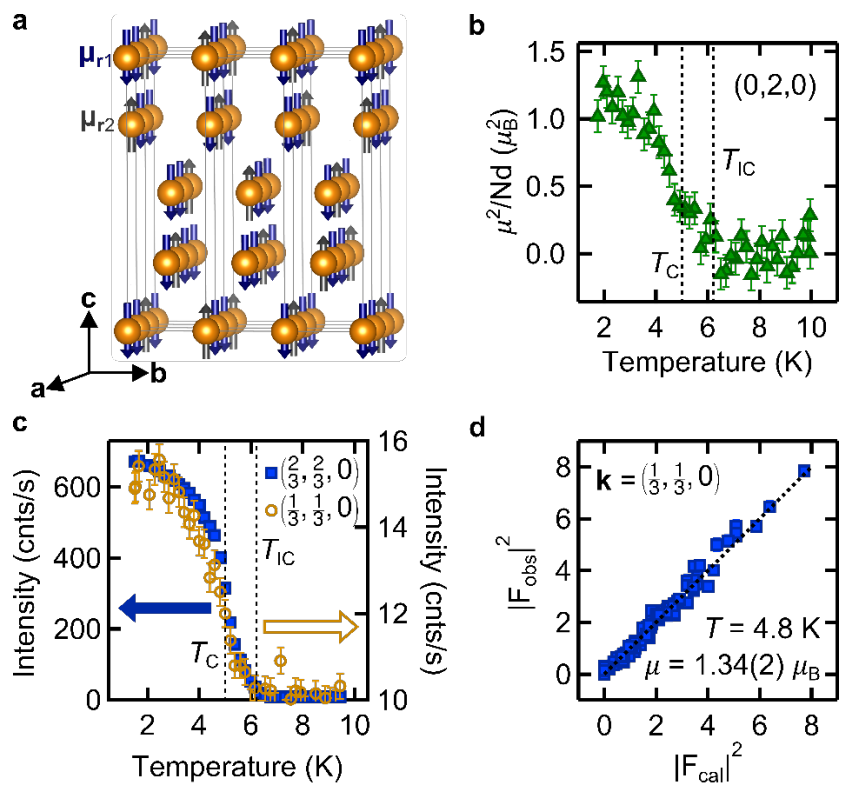

**Fig. 6**. (a) Schematic of the refined magnetic structure using the $Fd`d`2$ magnetic space group. The moments for the two Nd atoms within the magnetic unit cell are shown in blue ($\mu_{r1}$) and gray ($\mu_{r2}$). The non-magnetic atoms in the unit cell are omitted for clarity. (b) Order parameter for the

**k** = 0 FM component at **Q** = (0, 2, 0). The nuclear signal has been subtracted so all intensity is due to magnetism, and the *y*-axis represents the evolution of the net FM moment per Nd atom squared. (c) Order parameter for the $\left(\frac{2}{3},\frac{2}{3},0\right)$ (filled blue squares, left axis) and $\left(\frac{1}{3},\frac{1}{3},0\right)$ (open orange circles, right axis) magnetic Bragg peaks. (d) Rietveld refinement results from HB-3A showing the observed ($F_{obs}$) and calculated ($F_{cal}$) magnetic structure factors squared. Data are from the magnetic Bragg peaks taken at **G** + $\left(\frac{1}{3},\frac{1}{3},0\right)$ reflections at 4.8 K, where **G** are allowed nuclear Bragg peak reflections for a body centered tetragonal Bravais lattice. The black dashed line is $|F_{cal}|^2 = |F_{obs}|^2$.

## 3. f Incommensurate to Commensurate Transitions

Using the data from the WAND² experiment, we were able to resolve the small incommensurability associated with the magnetic structure between $T_{IC}$ and $T_C$. Data taken at $T_C$ < 5.5 K < $T_{IC}$ revealed magnetic Bragg peaks at positions **G** ± $\left(\frac{1}{3}-\delta,\frac{1}{3}-\delta,0\right)$ (where **G** are allowed nuclear Bragg peak reflections for a body centered tetragonal Bravais lattice) and Fig. 7 shows this incommensurability. Figs. 7a and b are an *H* and *K* cut of the data, respectively, through the (−1, 0, 1) Bragg peak position at 1.5 K, 5.5 K, and 9.0 K. Importantly, there is no observable shift in the peak position, although an increase in intensity is observed below 5.5 K as the **k**₀ magnetic component increases. Figs. 7c and d are similar *H* and *K* cuts, but through the magnetic Bragg peak stemming from the (−1, 0, 1) nuclear zone center, *i.e.* through (−1, 0, 1)−**k**₁. Here, a shift from the commensurate position of $(\delta, \delta, 0)$ is seen in the 5.5 K data and the wavevector is $\mathbf{k}_{IC} = \left(\frac{1}{3}-\delta,\frac{1}{3}-\delta,0\right)$. At this temperature, the shift is $\delta \approx 0.006$ r.l.u. Fig. 7 demonstrates the shift at one magnetic Bragg peak position, but it should be noted that this shift was observed for *all* magnetic Bragg peaks at 5.5 K when compared to 1.5 K.

A recent report [51] on the magnetic structure of NdAlGe utilized small-angle neutron scattering (SANS) to study the incommensurate to commensurate transition. Here, it was found that the appearance of a ferromagnetic component between $T_C$ <

$T < T_{IC}$ (just as in Fig. 6b) is, in fact, the third harmonic of the incommensurate SDW phase leading to magnetic peaks at positions $\pm(3\delta, 3\delta, 0)$ from nuclear zone centers. The intensity growth of the third harmonics indicates the SDW is squaring-up as well as locking-in to the commensurate phase at $T_C$. Here, the magnetic structure obtains its ground state as discussed in the previous section. As the data in Fig. 7 demonstrate, the incommensurability is small, and this detail could only be obtained from a SANS study as the incommensurability could not be resolved from the large nuclear Bragg peaks in a wide-angle diffraction experiment. Commonly, this type of incommensurate-to-commensurate squaring-up transition leads to a constant moment throughout the magnetic structure [52], which is the justification for constraining the moments to have equal magnitude in the ground state structure refinements.

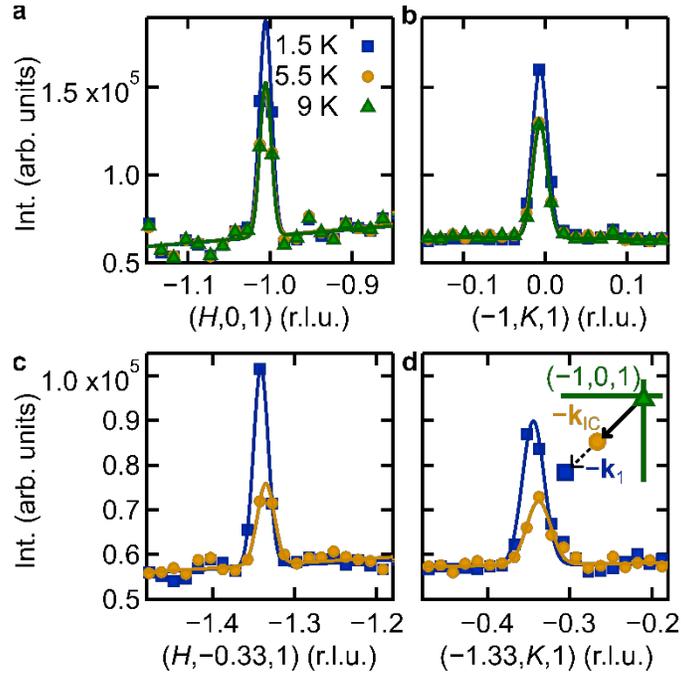

**Fig. 7.** Cuts through Bragg peaks from WAND$^2$ data. Integration about $L$ is ±0.4 r.l.u., and integration about $H$ (for cuts along $K$) and $K$ (for cuts along $H$) is ±0.1 r.l.u. (a) $H$ and (b) $K$ cuts taken through the (−1,0,1) Bragg peak at 9.0 K (green triangles), 5.5. K (blue squares), and 1.5 K (orange circles). These data show that no change in the peak position is observed as the temperature is lowered through $T_{IC}$ (6.3 K) and $T_C$ (5 K). (c) $H$ and (d) $K$ cuts through a magnetic Bragg peak

at 5.5 K (orange circles) and 1.5 K (blue squares). A shift of $(\delta, \delta, 0)$ from the commensurate $\left(-1\frac{1}{3}, -\frac{1}{3}, 1\right)$ position was observed in the 5.5 K data with $\delta = 0.006$ r.l.u. This shift was observed over all the antiferromagnetic magnetic Bragg peak positions within the detector range. The inset of (d) shows a schematic of the positions of the Bragg peaks shown in (a)-(d).

## 4. Electronic Structure Calculation

The NdAlGe cell was simulated in the fully polarized, or ferromagnetic (FM) state. Here, the Nd atoms are responsible for the magnetism of the structure, with the magnetic moment having negligible contributions from the Al and Ge atoms. In the FM configuration, each Nd atom presents a magnetic moment equal to 2.96 $\mu_B$. This value is similar to the average magnetic moment measured experimentally at 3.03 $\mu_B$. The antiferromagnetic (AFM) and ferrimagnetic (FiM)-duu configurations are contrastingly more complex. Due to the complexity of both models, we only evaluated the FM structure. The FiM structure is characterized by a total magnetic moment different from zero. Similar results are expected if we apply the same analysis as the one reported here for the FM magnetic structure.

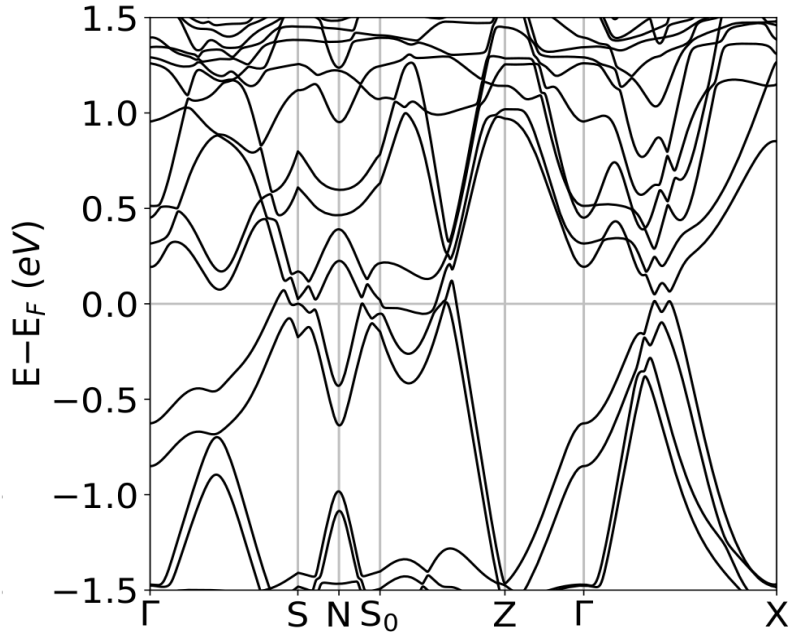

**Fig. 8.** Band diagram of the NdAlGe FM structure under the DFT+SOC+U framework.

Fig. 8 shows the band structure of the FM NdAlGe structure. The FM behavior is corroborated by noticing that there is always a significant gap between the bands that have the same slope, belonging to electrons with different spin states. It is worth mentioning that the addition of the +U functional pushes the region of high density of states away from the Fermi level leaving only a small number of Fermi surface band crossings. Additionally, Weyl points are clearly present in the trajectories Γ-S and Γ-X close to the Fermi level. This is further discussed below.

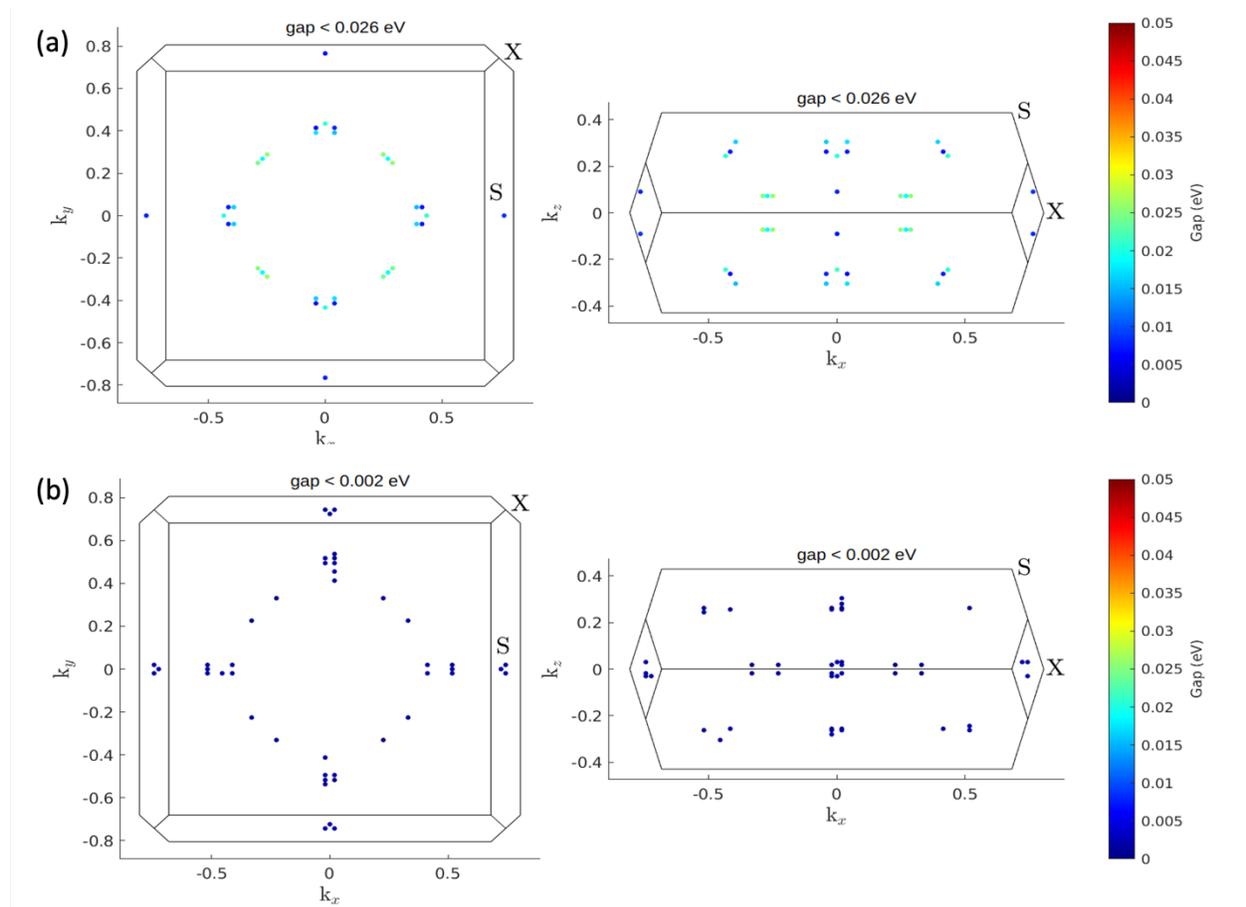

**Fig. 9.** Weyl nodes in the top (left) and front (right) views of the Brillouin zone with their respective energy gaps of DFT+U FM NdAlGe. Subfigures with a gap smaller than (a) 0.026 eV restrict the sign of the valence and conduction energies. Subfigure (b) removes this restriction and considers a gap smaller than 0.002 eV.

The determination of Weyl points is qualitatively easy. However, due to the approximate nature of the *ab*-initio calculations, a compromise between computational precision and physical exactitude must be made. More specifically, the quantitative determination of Weyl points in our calculations was done considering only the closest nodal points to the Fermi surface. To this end, the nodal points must have negative valence energy, positive conduction energy, and a gap smaller than a certain criterion between the two energies.

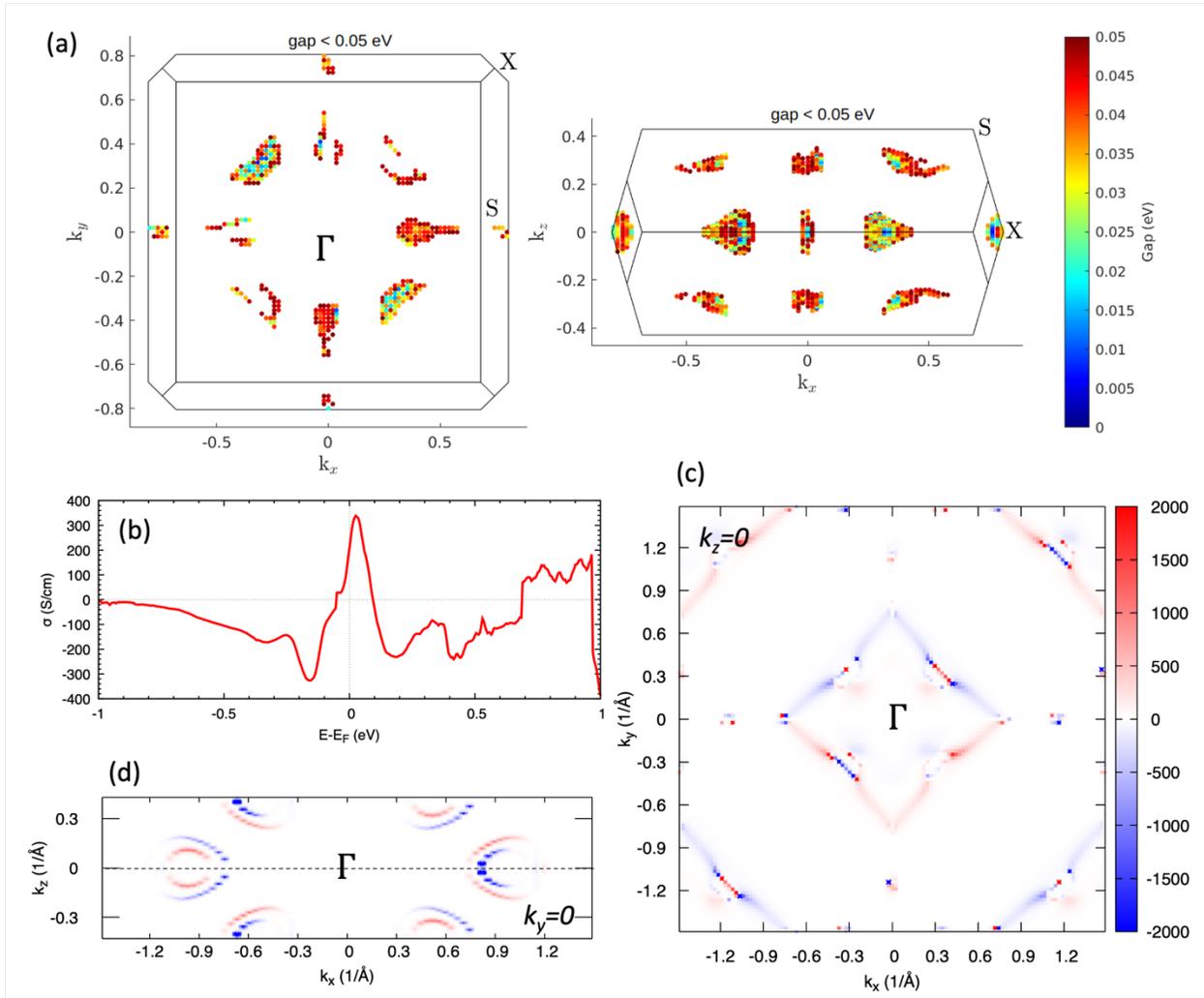

**Fig.10.** Weyl nodes in the top (left) and front (right) views with their respective energy gap from DFT+SOC+U FM NdAlGe. Subfigure (a) 0.05 eV without valence and conduction band restriction, (b) magnitude of the Anomalous Hall Conductivity as a function of energy. The Fermi energy is indicated by the dotted vertical line at zero energy, and the Berry curvature at 1 emu/(mol Oe) = $4\pi \; 10^{-6}$ m$^3$/mol. (c) $k_z = 0$ and (d) $k_y = 0$ planes depicting Weyl nodes as red/blue peaks at the Berry curvature.

In the DFT+U calculations, there are 48 Weyl points with a gap smaller than 0.019 eV and 70 Weyl points with a gap smaller than 0.026 eV. The former is the minimum gap in which Weyl points are observed along the Γ-X direction. See Figure 9a for a graphical representation of the points for a gap smaller than 0.026 eV in the Brillouin zone. If we remove the restriction on the signs of the band energies of the nodal points, then the gap is reduced to 0.002 eV. In this situation, there are a total of 72 Weyl points in the Brillouin zone (see Figure 9b). In all cases, the Weyl points are located around the Γ-S and Γ-X directions and are symmetric around Γ.

The addition of the SOC term in the Hamiltonian perturbs the system and prevents a quantitatively precise determination of the Weyl points. A greater value for the energy gap is necessary to obtain nodal points in the directions of interest. By considering a gap of 0.025 eV, there are 43 Weyl points in the Brillouin zone. By removing the restriction in the signs of the band energies, there are 72 Weyl points with a gap of 0.01 eV. Upon considering the highest threshold for the gap (0.05 eV, Figure 10a) the distribution of the Weyl points in the Brillouin zone begins to look like nodal surfaces and resemble the form of the Fermi surface.

We further calculate the anomalous Hall conductivity (Figure 10b) for the FM system setting the Fermi energy as reference. Our calculated value for $\sigma \approx 270\ \Omega^{-1}\mathrm{cm}^{-1}$ is in reasonable agreement with our experimental value ($\approx 430\ \Omega^{-1}\mathrm{cm}^{-1}$) in both order of magnitude and sign. As mentioned before, the large Anomalous Hall Conductivity is intrinsic due to the Berry curvature most likely generated by the Weyl points. To prove if the Weyl nodes generate such Berry curvature, we plot it at $k$-planes at the Weyl point positions. Figs. 10b and c depict the Berry curvature for $k_z = 0$ and for $k_y = 0$. These figures demonstrate that the Weyl points generate Berry curvature, which drives the intrinsic anomalous Hall effect already measured and confirmed by our electronic structure calculations. This effect likely appears because magnetism in Weyl semimetals modifies the Weyl nodes to induce a Berry curvature field, further generating several interesting phenomena such as the observed Anomalous Hall Effect.

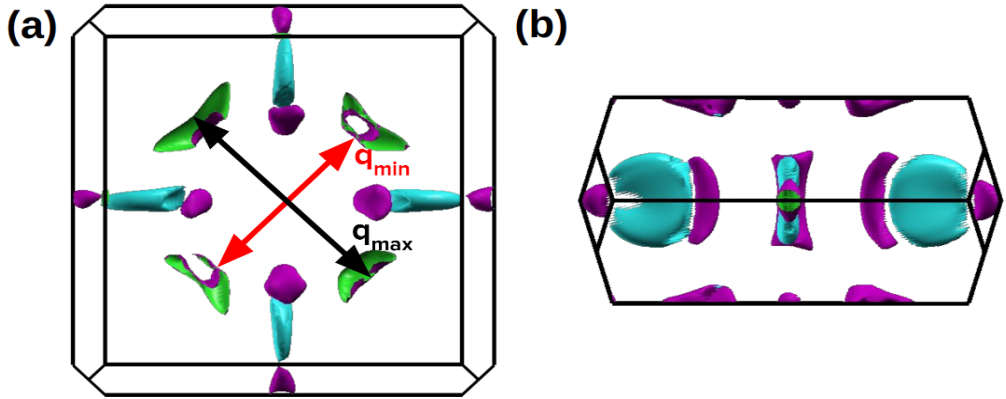

**Fig. 11.** Fermi surface and nesting vector in the DFT+SOC+U FM NdAlGe. (a) Shows a top view and (b) a front one.

Finally, Fig. 11 displays the Fermi surface of NdAlGe. Within this surface, we notice that the regions displayed in green in Fig. 11a are parallel to each other over a significant area. These are connected by the red and black nesting vectors drawn along the [110] and [1-10] directions. The lengths of the nesting vectors $q_{min}$ and $q_{max}$ are 0.63 Å$^{-1}$ and 0.80 Å$^{-1}$, respectively.

## 5. Discussion and Conclusion

The two thermodynamic transitions observed in the resistivity, magnetic susceptibility, and heat capacity are related to the change in magnetic state from paramagnetic ($T > T_{IC}$) to incommensurate spin density wave state for $T_C \leq T \leq T_{IC}$ and to a commensurate ferrimagnetic state at $T \leq T_C$. The presence of both itinerant and local moments is inferred from the ratio of the effective moment ($\mu_{eff}$) from Curie-Weiss law and the saturation magnetization ($M_S$) obtained from magnetization measurements at high magnetic fields and low temperatures. This ratio for NdAlGe is $\frac{\mu_{eff}}{M_S} = 1.2 > 1$, suggesting that the itinerant moments reduce the net saturation magnetization. The importance of the itinerant conducting carriers in determining the magnetism of NdAlGe is further supported by our electronic structure calculations where the calculated Fermi surface, shown in Fig. 11a, include small hole pockets (green) near $Q = \pm (\frac{1}{3}, \frac{1}{3}, l)$ that display well-nested surfaces. Typically, incommensurate spin density wave order in local moment systems, such as those involving rare earth ions, occurs via the RKKY interaction [53–55]. However, in a system such as NdAlGe the nested Fermi surfaces may affect the RKKY interactions between local moments creating an incommensurate spin density wave. In such a case, the magnetic wavevector for both the itinerant and local moments is set by the Fermi surface nesting condition. Such a scenario has been suggested in other rare earth compounds such as NdAlSi [8] and GdSi [54]. Since, in the case of NdAlGe, the nested pockets are Weyl like, and since the nesting vector determines the magnetic ordering vector, we conclude that the magnetism in NdAlGe is mediated by the itinerant Weyl fermions in a manner similar to that found in NdAlSi [8].

The magnetoresistance in NdAlGe also displays interesting behavior. As shown in Fig. 4 the magnetoresistance increases with field in the commensurate (uud) type-ferrimagnetic state (Region I), decreases in the transition region between I and II, and then reverses course with further increasing field having a positive MR in region II where the magnetic moments are in an (uuu) state. The magnetoresistance is large and increases by 15% for $H // c$ at $T = 0.4$ K and $H = 180$ kOe. The positive variation of magnetoresistance in the field polarized state is rather counterintuitive and may reflect a change in the electronic structure and Fermi surface topology as in other rare earth compounds such as NdAlSi [44] and heavy fermion metals [45,56]. Unfortunately, the disorder present in our sample (RRR ≈ 2.5) prevents us from observing the quantum oscillations to further test this hypothesis. Future field-dependent magnetic and electronic structure analyses are required to fully understand the positive magnetoresistance in NdAlGe.

Our investigation of the Hall effect in NdAlGe revealed two unusual anomalous responses in the magnetic state. The anomalous Hall conductivity is surprisingly large (430 $\Omega^{-1}$ cm$^{-1}$, $T$ = 2 K, region I) and (1030 $\Omega^{-1}$ cm$^{-1}$, $T$ = 2 K, region II). Such a large anomalous Hall conductivity cannot arise from an extrinsic skew scattering or side jump mechanism in NdAlGe with the conductivity lying within the moderately disordered range. These extrinsic contributions can give such large anomalous Hall conductivity only in the ultraclean limit [57–59] which is not the case in current sample. In fact, the obtained anomalous Hall conductivity in NdAlGe is of the same order of magnitude as the calculated intrinsic Berry curvature contribution from electronic structure and comparable to other members of this family [10,13]. Further, the presence of Weyl nodes near the Fermi surface and the large anomalous Hall response comparable to or greater than the intrinsic Berry curvature contribution limit indicates that the anomalous Hall effect in NdAlGe is caused by the large Berry curvature associated with the Weyl Fermions [47,60]. The two slightly different values of $Rs$ are an indication of the electronic structure (and associated Berry curvature) modification with the field. This picture of electronic structure modification is also consistent with the positive magnetoresistance in region II. The large magnitude of the intrinsic anomalous Hall response which is related to the topology of electronic structure has also observed in other rare earth compounds such as DyPtBi [61], TbPtBi [62] and PrAlGe [13] giving us further confidence in our conclusion.

We also investigated the possibility of the presence of topological and anomalous (Berry phase related) Hall effects. In this scenario, there is only one anomalous Hall coefficient ($R_S^{II}$) which can be obtained by fitting the high field (region II) data in

Fig. 5b. Any low field data (region I), that cannot be fitted with this large field anomalous Hall coefficient is the topological Hall effect ($\rho_{xy}^T$) arising from the non-trivial topology of magnetic structure in region I. Assuming such scenario, the variation of $\frac{\rho_{xy}^T}{H} = \frac{\rho_{xy}}{H} - R_0 - 4\pi R_S^{II} \frac{M}{H}$ is presented in Fig. 12. From Fig. 12, it appears that there is a large topological Hall effect in low field regime and the topological component disappears only in region II. The generation of such a large topological Hall effect requires a highly non-collinear and non-coplanar magnetic structure. Our neutron diffraction experiments indicate an almost collinear (along c-axis) magnetic structure with only small ab-plane components. Such a magnetic structure is very unlikely to generate such a large topological Hall response.

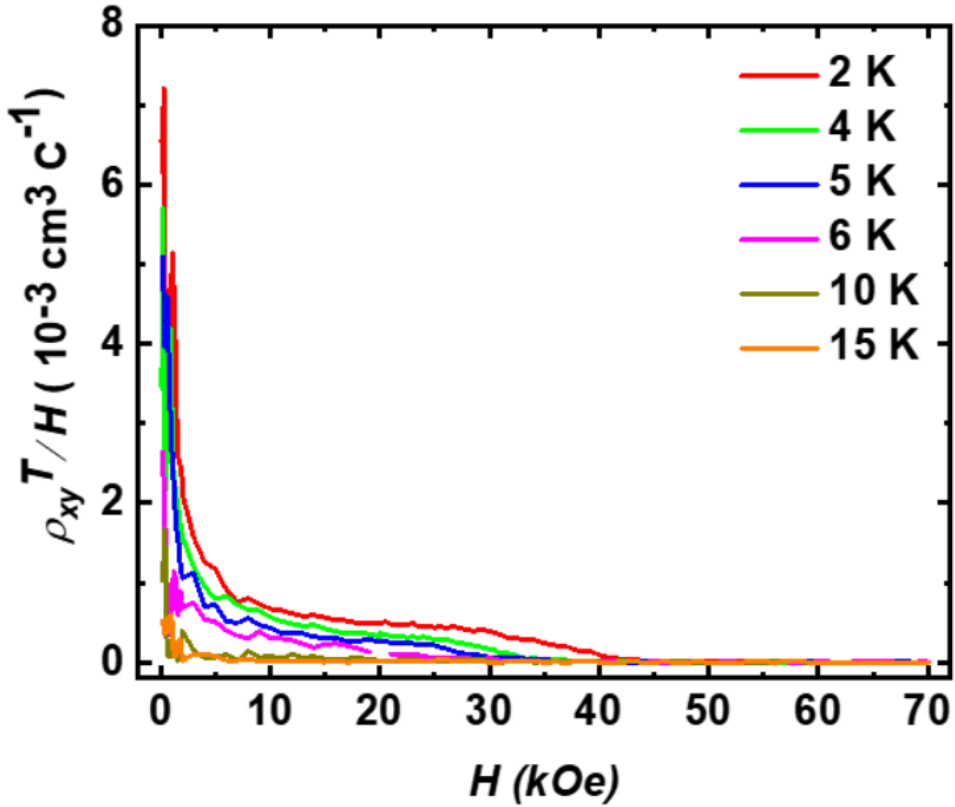

**Fig. 12**: Variation of $\frac{\rho_{xy}^T}{H}$ with $H$ for NdAlGe at select temperatures.

In summary, our results indicate that the incommensurate spin density wave order and the large anomalous Hall response in NdAlGe are directly related to the nested Fermi surfaces containing Weyl nodes. The appearance of a multi-$k$ structure along with a large anomalous Hall response is similar to what has been found in other compounds of this family. This establishes that the RAlX family of materials are good candidates to investigate the emergent electronic and magnetic properties arising from the interplay of itinerant Weyl fermions and local magnetic moments and the possibility of controlling them through application of magnetic or electronic fields. Our work is pivotal in revealing the participation of relativistic fermions in controlling the collective behavior of materials such as magnetism. Future small-angle neutron scattering, and magnetic field-dependent neutron diffraction measurements are desired to understand the possibility of long-period topological magnetic phases and their evolution with magnetic fields and temperature.

**Note:** During the preparation and review process of this manuscript similar results were reported a recent work [51] providing independent verification of our results.

**Acknowledgments:**

C. D acknowledges the helpful discussion with M. Asmar. This work is based upon the work supported by National Science Foundation under grant number DMR-2213443. The neutron diffraction experiments used resources at the High Flux Isotope Reactor, a DOE Office of Science User Facility operated by the Oak Ridge National Laboratory. A portion of this work was performed at the National High Magnetic Field Laboratory, which is supported by the National Science Foundation Cooperative Agreement No. DMR-1644779 and the State of Florida. The identification of any commercial product or trade name does not imply endorsement or recommendation by the National Institute of Standards and Technology. R.C and R. J are supported by the U.S. Department of Energy grant number DE- SC0012432. R.R, R.G and J.G acknowledge DGAPA-UNAM project IA100822 for partial financial support. Calculations were performed in the DGCTIC-UNAM

Supercomputing Center project LANCADUNAM-DGTIC-368. JGS acknowledges A. Rodriguez-Guerrero for the technical support.

**Appendix A: Crystal structure determination using single crystal and powder neutron diffraction.**

Single crystal neutron diffraction on HB-3A in four-circle mode was used to determine the crystal structure at 4.8 K. A total of 127 nuclear peaks were used to solve the nuclear structure which was determined to be the same as at room temperature. The lattice parameters at 4.8 K were found to be $a = b = 4.224$ Å and $c = 14.624$ Å. The non-centrosymmetric structure and the variation of observed and calculated structure factors are shown in Fig. 13.

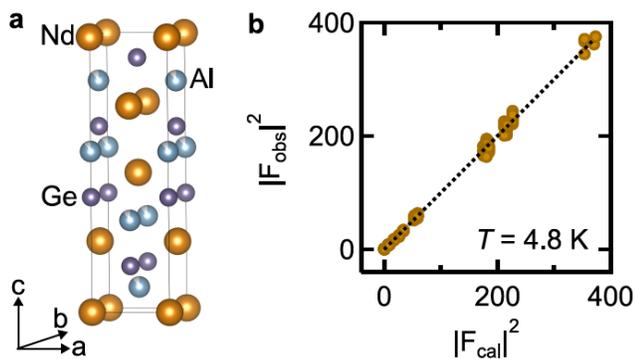

**Fig. 13:** (a) NdAlGe crystal structure with space group, $I4_1md$ (#109). (b) Rietveld refinement results for the nuclear structure taken at 4.8 K. Data are from single crystal neutron diffraction on HB-3A. Goodness of fit parameters can be found in Table I of the main text.

In addition to single crystal neutron diffraction, we also performed neutron powder diffraction on the polycrystalline sample prepared by arc melting and subsequent vacuum annealing at 1000 $^0$C for 4 days. This is to ensure that this compound forms in non-centrosymmetric crystal structure despite slight variations in stoichiometry. We refined the same neutron data using both centrosymmetric ($I4_1/amd$) and non-centrosymmetric ($I4_1md$) space groups. The quality of refinement is much better with non-centrosymmetric structure.

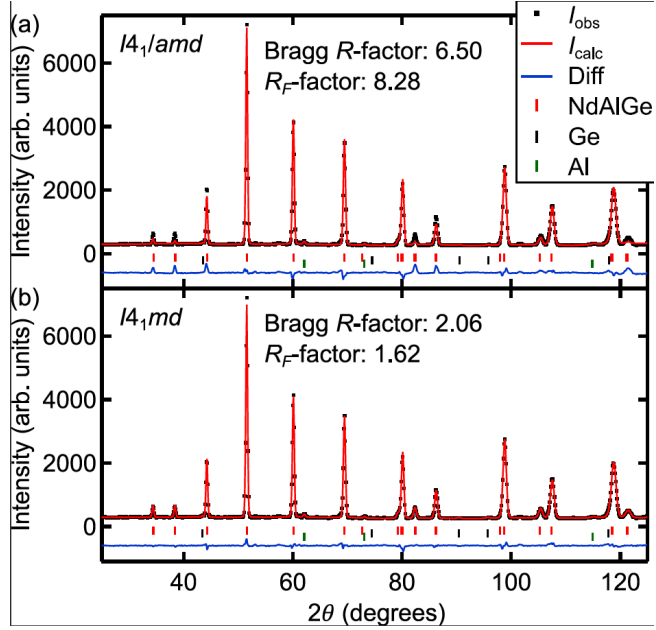

**Fig. 14** Rietveld refinement of powder neutron pattern of NdAlGe at 20 K. (a) Using centrosymmetric (I4$_1$/amd) and, (b) using non-centrosymmetric (I4$_1$md) space groups. Goodness of fit parameters can be found in Table I of the main text.

## Appendix B: Resistivity measurements

The resistivity was measured using four platinum electrodes mounted on the sample using silver epoxy. The current was applied along *a* axis.

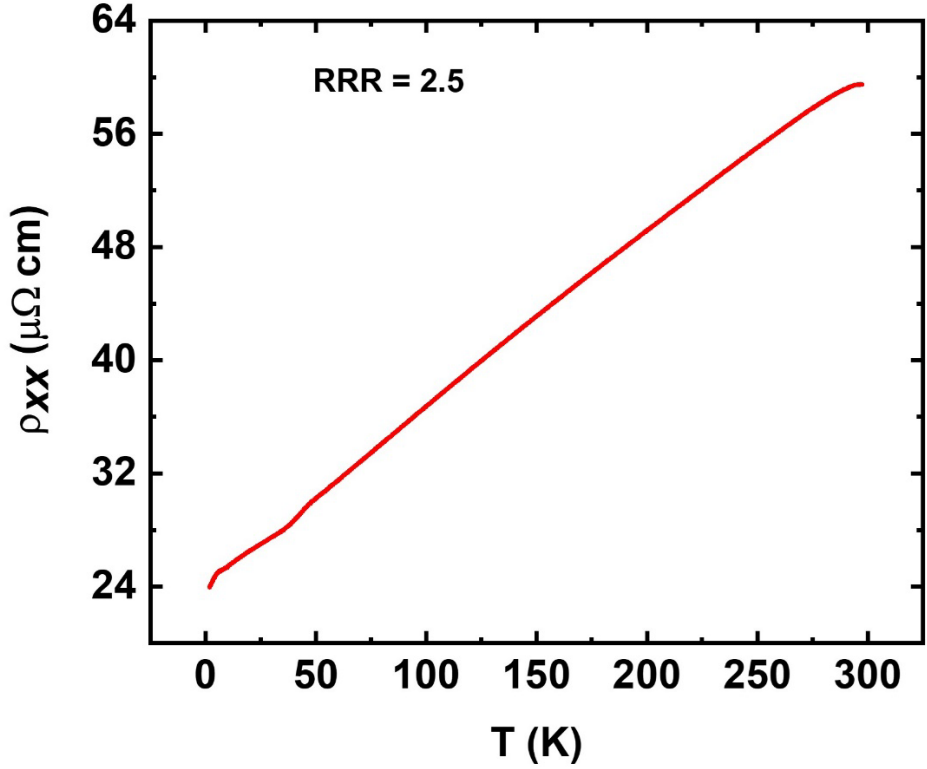

**Fig. 15** Variation of longitudinal resistivity, $\rho_{xx}$, with temperature, $T$ for NdAlGe.

## Appendix C: Estimation of anomalous Hall Conductivity

The anomalous Hall conductivity ($\sigma_{xy}^A$) is given by relation: $\sigma_{xy}^A = \frac{\rho_{xy}^A}{(\rho_{xy}^A)^2 + (\rho_{xx})^2}$. Here $\rho_{xy}^A$ is the anomalous Hall resistivity, and $\rho_{xx}$ is the linear resistivity. The anomalous Hall resistivity ($\rho_{xy}^A$) is related to the Hall resistivity $\rho_{xy}$ by a relation $\rho_{xy} = R_0 H + \rho_{xy}^A$. Now $\rho_{xx}$ is taken from the data presented in Fig 15. To find $\rho_{xy}^A$, we plotted $\rho_{xy}$ vs $H$ as shown in Fig 5a and Fig. 16. As can be seen in Fig. 16, the variation of $\rho_{xy}$ vs $H$ can be fitted with two straight lines [olive (region I) and violet (region II)]. The intercepts of these straight lines give anomalous Hall resistivity in region I and II. To take an example at $T = 2$ K, $\rho_{xx}$ ($H = 0$) =23.99 µΩ cm, $\rho_{xy}^{A,I}$ (region I intercept)= - 0.247 µΩ cm, and $\rho_{xy}^{A,II}$ (region II intercept) = - 0.596 µΩ cm. This gives $|\sigma_{xy}^{A,I}| = |\frac{\rho_{xy}^{A,I}}{(\rho_{xy}^{A,I})^2 + (\rho_{xx})^2}| = 430$ Ω$^{-1}$ cm$^{-1}$ and $|\sigma_{xy}^{A,II}| = |\frac{\rho_{xy}^{A,II}}{(\rho_{xy}^{A,II})^2 + (\rho_{xx})^2}| = 1034$ Ω$^{-1}$cm$^{-1}$.

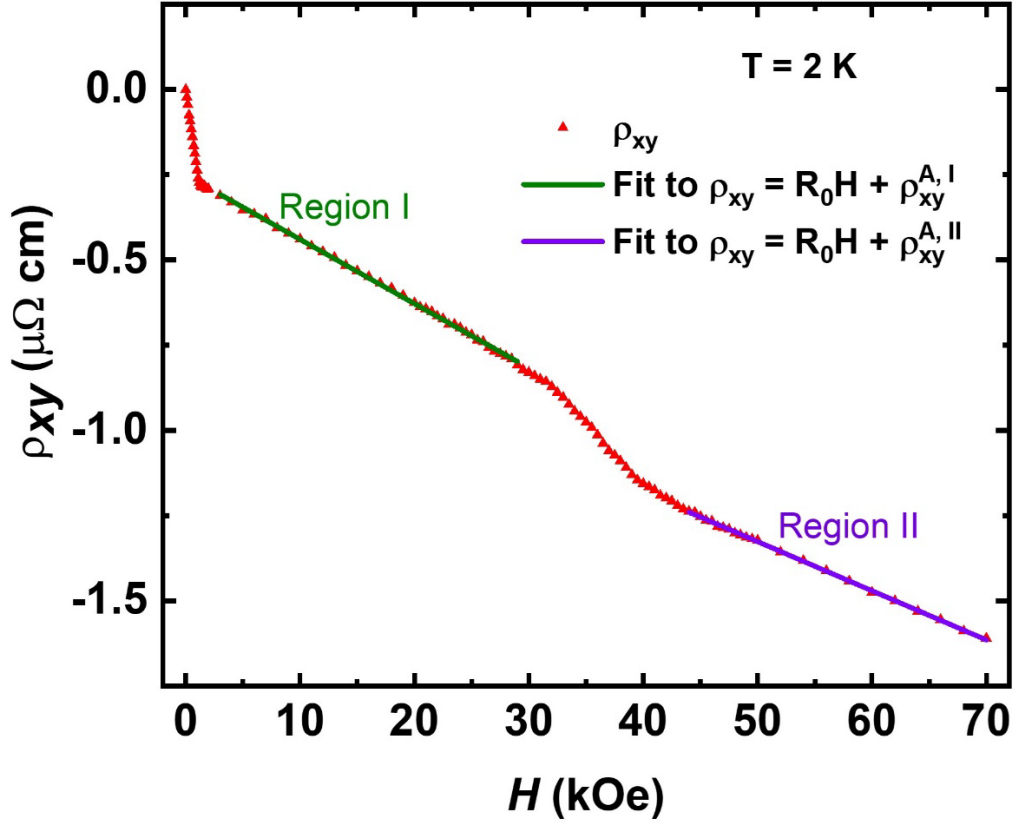

**Fig. 16** Variation of Hall resistivity ($\rho_{xy}$) with magnetic field ($H$) for NdAlGe at $T = 2$ K. The data is fitted with two straight lines with different intercepts in region I and region II as shown in Figure. These two intercepts result in two different anomalous Hall conductivities.

## Appendix D: Magnetic Structure Details

The crystallographic symmetry operators in the paramagnetic phase are those for the space group $I4_1md$ and belong to a group, $G_0$. The symmetry elements which keep **k** invariant belong to a coset known as the "little group", $G_\mathbf{k}$. There exist other cosets which lead to inequivalent **k**-vectors called "arms-of-the-star", which for NdAlGe are $\mathbf{k}_1 = \left(\frac{1}{3}, \frac{1}{3}, 0\right)$, $\mathbf{k}_2 = \left(-\frac{1}{3}, -\frac{1}{3}, 0\right)$, $\mathbf{k}_3 = \left(-\frac{1}{3}, \frac{1}{3}, 0\right)$, and $\mathbf{k}_4 = \left(\frac{1}{3}, -\frac{1}{3}, 0\right)$. In a single-**k** magnetic structure with multiple **k**-vectors in the arms-of-the-star, there exist domains (called **k**-domains) which form energetically equivalent magnetic structures that are macroscopically separated within the sample. Each **k**-domain contributes to a different set of magnetic Bragg peaks stemming from a Brillouin zone center. If **k** and -**k** are inequivalent both must participate in the modulation for a single domain. This is the case for the NdAlGe magnetic structure described above and so, there would exist a second domain described by $\pm\mathbf{k}_3 = \pm\left(-\frac{1}{3}, \frac{1}{3}, 0\right)$ and $\mathbf{k}_0 = 0$; however, a multi-**k** structure which includes all arms of the $\mathbf{k}_1$ star (and $\mathbf{k}_0 = 0$) should not be excluded. A multi-**k** structure may exist, especially when crystal electric fields and/or higher order exchange play a role in the magnetic structure [52] Here, the magnetic Bragg peak satellites stemming from nuclear zone centers are due to a single magnetic domain where all arms-of-the star are participating in forming the magnetic structure. Magnetic Bragg peaks at $\pm\mathbf{k}_3$ positions from Brillouin zone centers are observed in NdAlGe, however neutron diffraction cannot distinguish between multi-**k** and **k**-domain structures without the application of an external perturbation [52], such as a magnetic field or uniaxial pressure, which would break **k**-domain degeneracies should they exist.

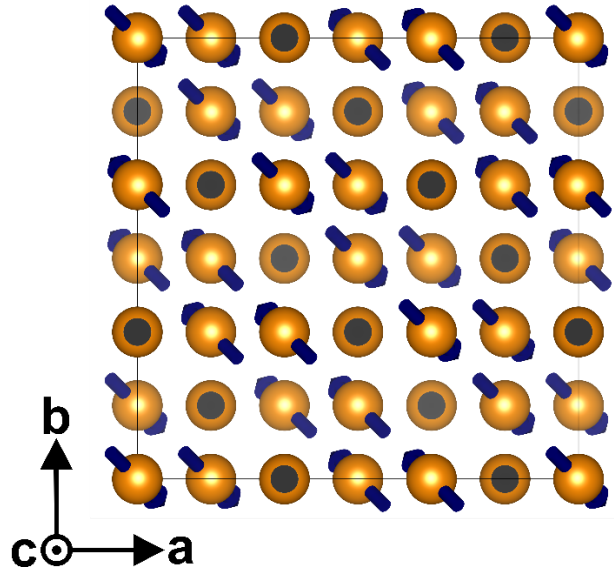

**Fig. 17.** The *ab*-plane view of the NdAlGe (3*a*, 3*b*, *c*) magnetic unit cell. Magnetic space group $Fd`d`2$ (#43.227) contains two Nd sites in the (3*a*, 3*b*, *c*) unit cell: $\mathbf{r}_1 = (0,0,0)$ (shown as the blue moments) and $\mathbf{r}_2 = \left(\frac{1}{6}, \frac{1}{6}, \frac{1}{2}\right)$ (shown as the gray moments which point along the +*c*-axis). An *ab*-plane component is allowed for $\mathbf{r}_1$, where the moment can point in any direction within this plane. Using the NdAlSi canting motif, where the moments are along the [-1,1,0] direction, we can estimate the in-plane moment using the intensity collected at the following Bragg peak positions (with respect to the nuclear unit cell): $\pm\left(\frac{1}{3}, \frac{1}{3}, 0\right), \left(-\frac{1}{3}, 1\frac{2}{3}, 0\right), \left(1\frac{2}{3}, -\frac{1}{3}, 0\right), \left(\frac{1}{3}, 2\frac{1}{3}, 0\right),$

$\left(1\frac{2}{3}, -2\frac{1}{3}, 0\right)$. The *c*-axis magnetic structure does not contribute any intensity at these positions and the fit leads to a moment of 0.22(1) $\mu_B$. The canting structure shown above exaggerates in the in-plane moment for clarity.

**Appendix E: Electronic structure calculations**

The links for videos showing the appearance of nodal points with the band energies and gap criteria on the DFT+U and DFT+SOC+U are shown in this section:

DFT+U: https://drive.google.com/drive/folders/1TOjWP_mRT_jQ2aqg8s-QSU0Vik4sGmiO?usp=sharing

DFT+SOC+U:

https://drive.google.com/drive/folders/1L1NBknRS32ZyVCSG75KmTyrDj_d6jhYj?usp=sharing